# Synthesis of new high-entropy alloy-type Nb$_3$ (Al, Sn, Ge, Ga, Si) superconductors

Aichi Yamashita[1], Tatsuma D. Matsuda[1], Yoshikazu Mizuguchi[1,] *

1. Department of Physics, Tokyo Metropolitan University, 1-1, Minami-osawa, Hachioji, 192-0397.

*Corresponding author: Yoshikazu Mizuguchi.

E-mail: mizugu@tmu.ac.jp

**Abstract**

Studies on high-entropy alloy (HEA) superconductors have recently been increasing, particularly in the fields of materials science and condensed matter physics. To contribute to research on new HEA-type superconductors, in our study we synthesized polycrystalline samples of A15-type superconductors of Nb$_3$Al$_{0.2}$Sn$_{0.2}$Ge$_{0.2}$Ga$_{0.2}$Si$_{0.2}$ (#1) and Nb$_3$Al$_{0.3}$Sn$_{0.3}$Ge$_{0.2}$Ga$_{0.1}$Si$_{0.1}$ (#2) with an HEA-type site by arc melting. Elemental and structural analyses revealed that the compositions of the obtained samples satisfied the HEA state criteria. Superconducting transitions were observed at 9.0 and 11.0 K for #1 and #2, respectively, in the temperature dependence of magnetization and electrical resistivity. Specific heat measurements revealed that the Sommerfeld coefficient, Debye temperature, and $\Delta C/\gamma T_c$ for the obtained samples were close to those reported for conventional Nb$_3$Sn family superconductors.

Highlights

・New high-entropy alloy (HEA)-type Nb$_3$$M$ superconductors ($M$: Al, Sn, Ge, Ga, Si) were synthesized.

・A superconducting transition was observed for Nb$_3$Al$_{0.3}$Sn$_{0.3}$Ge$_{0.2}$Ga$_{0.1}$Si$_{0.1}$ at 11.0 K.

・The HEA state at the $M$ site of Nb$_3M$ resulted in a lower $T_c$.

**1. Introduction**

High-entropy alloys (HEAs) are typically defined as alloys containing at least five elements with concentrations between 5 and 35 at. % [1, 2], resulting in high configurational mixing entropy ($\Delta S_{mix}$), defined as $\Delta S_{mix} = -R\Sigma_i c_i \ln c_i$, where $c_i$ and $R$ are the compositional ratio and the gas constants, respectively [2]. HEAs have recently attracted much attention in the fields of materials science and engineering because of their tunable properties as structural materials, such as excellent mechanical performance under extreme conditions [1, 2]. As HEA superconductors, simple alloys with bcc, α-Mn, CsCl, and hcp crystal structures have mainly been studied so far [3–14]. Among these, Ta$_{0.34}$Nb$_{0.33}$Hf$_{0.08}$Zr$_{0.14}$Ti$_{0.11}$ with a superconducting transition temperature ($T_c$) of 7.3 K, exhibits robustness in the superconducting state under extremely high pressures up to 190 GPa [4]. This suggests that HEA superconductors can be applied under extreme conditions and the study of HEA superconductors can be accelerated.

Thus far, we have extended the concept of HEA to compounds [15–22], wherein one of the crystallographic sites is high-entropy alloyed. Recently, we reported layered superconductors with HEA-type crystallographic sites: HEA-type cuprate $RE$123 (RE: rare earth elements) and $RE$O$_{0.5}$F$_{0.5}$BiS$_2$ with a HEA-type rare earth (RE) site [15–17]. In a RE123-type cuprate, no degradation of $T_c$ was confirmed by increasing $\Delta S_{mix}$ [15]. In addition, the emergence of bulk superconductivity has been observed in HEA-type $RE$O$_{0.5}$F$_{0.5}$BiS$_2$ [16, 17]. Notably, crystal structural analysis revealed suppression of the local structural disorder, which corresponds to the improvement of the bulk nature of superconductivity in this system, with an increase in $\Delta S_{mix}$ [17]. Based on these findings, the HEA effects in layered superconductors seem to work positively or at least less negatively. To investigate the effect of HEA on non-layered compounds, we also investigated NaCl-type metal chalcogenide superconductors with high $\Delta S_{mix}$ [20–22]. For instance, an HEA-type AgInSnPbBiTe$_5$ superconductor with $T_c$ = 2.6 K was reported. In contrast to layered systems, NaCl-type HEA tellurides exhibited lower $T_c$ than low-entropy tellurides [21]. These results suggest that the effects of HEAs on the superconducting properties of compounds depend on their crystal structure and dimensionality.

To further investigate the effects of HEAs on superconducting properties, we focused on A15-type Nb$_3M$ ($M$: Al, Sn, Ge, Ga, Si) compounds, which have a relatively high $T_c$ (above 18 K) and upper critical fields ($H_{c2}(0)$) of approximately 30 T [23–33]. They are well known as practical materials for superconducting magnets higher than 10 T [33]. In this paper, we report on the syntheses and properties of new HEA-type Nb$_3M$ superconductors, wherein the composition at the $M$ site satisfies the compositional criterion of the HEA and achieves $\Delta S_{mix}$ above $1.5R$.

## 2. Experimental details

Polycrystalline samples of Nb$_3$Al$_{0.2}$Sn$_{0.2}$Ge$_{0.2}$Ga$_{0.2}$Si$_{0.2}$ (#1) and Nb$_3$Al$_{0.3}$Sn$_{0.3}$Ge$_{0.2}$Ga$_{0.1}$Si$_{0.1}$ (#2) were synthesized by arc melting in an Ar atmosphere. Pure metal powders of Nb (99.9%), Al (99.9%), Sn (99.9%), Ge (99.99%), and Si (99.999%) were mixed with the above-mentioned composition and pelletized. Metal pellets and pure Ga (99.9999%) grains were used as starting materials for the arc melting process. Arc melting was repeated after turning the sample three times to ensure homogenization of the sample. The obtained samples were characterized using energy-dispersive X-ray fluorescence (XRF) analysis on a JSX-1000S instrument (JEOL). The phase purity and crystal structure were examined through X-ray diffraction (XRD) with Cu-K$\alpha$ radiation on a Miniflex-600 (RIGAKU) equipped with a high-resolution semiconductor detector D/tex-Ultra. The obtained XRD patterns were refined using the Rietveld method and RIETAN-FP [34]. A schematic image of the refined crystal structure was produced using VESTA [35]. To investigate the superconducting properties of the samples, magnetization was measured using a superconducting quantum interference device (SQUID) on a magnetic property measurement System-3 (MPMS-3, Quantum Design). The temperature dependence of magnetization was measured after zero-field cooling (ZFC) and field cooling (FC). The magnetic field dependence of magnetization was measured from -7 to 7 T at 2.0 and 4.2 K. Temperature dependence of electrical resistivity under magnetic fields up to 9 T was measured using a physical property measurement system (PPMS, Quantum Design) via a conventional four-probe method. The temperature dependence of specific heat was measured by a PPMS using a relaxation method under various fields of 0, 2, 6, and 9 T.

## 3. Results and discussion

The powder XRD patterns of $Nb_3Al_{0.2}Sn_{0.2}Ge_{0.2}Ga_{0.2}Si_{0.2}$ (#1) and $Nb_3Al_{0.3}Sn_{0.3}Ge_{0.2}Ga_{0.1}Si_{0.1}$ (#2) are shown in Figure 1. The XRD peaks of the major phase can be indexed using the cubic $Cr_3Si$-type ($Pm\bar{3}m$) model. The lattice constant was estimated using Rietveld refinement and is plotted in Table 1. A schematic image of the refined crystal structure of sample #2 is shown in the inset of Figure 1. Impurity phases of $Nb_5Si_3$ were detected at 8.5% and 4.5% for samples #1 and #2, respectively. $Nb_5Si_3$ is non-superconducting at $T > 1$ K. The compositions of #1 and #2 were estimated to be $Nb_{3.182}(Al_{0.131}Sn_{0.172}Ge_{0.213}Ga_{0.175}Si_{0.128})$ and $Nb_{3.242}(Al_{0.170}Sn_{0.228}Ge_{0.220}Ga_{0.069}Si_{0.071})$ by XRF measurements. The calculated $\Delta S_{mix}$ for M site was $1.59R$ and $1.50R$ for #1 and #2, respectively. Note that, we assumed the total molar ratio of $M$ ($M$ = Al, Sn, Ge, Ga, Si) site is 1.0.

Figures 2(a) and 2(b) show the temperature dependences of magnetization for #1 and #2 that were measured after ZFC and FC at a magnetic field of 10 Oe. Large diamagnetic signals corresponding to the appearance of superconductivity are observed below 9.0 and 11.0 K for #1 and #2, respectively. The estimated shielding volume fraction from the ZFC values at 2.0 K exceeded 100% shielding. A typical $M$-$H$ loop was observed for samples #1 and #2 at 2.0 K. The $\Delta M$ at each magnetic field, which is related to the critical current density ($J_c$), for #2 is slightly larger than that for #1.

Figure 3(a) shows the temperature dependence of the electrical resistivity ($\rho$) for #1 and #2. The temperature dependences of $\rho$ for #1 and #2 exhibited metallic behaviors with a very small residual resistivity ratio ($RRR = \rho_{300K}/\rho_{13K}$) of approximately 1.2 for both samples, which is often observed in conventional HEA superconductors [3–5, 20–22]. Compared to the superconducting transition of sample #1, sample #2 exhibited a broader transition. The transition width ($\Delta T_c$) of #1 and #2 is estimated as 0.14 and 0.75 K, respectively.

Figures 3(b) and 3(c) show the temperature dependence of resistivity under various magnetic fields ($H$ = 0, 1, 2, 3, 5, 7, and 9 T) near the superconducting transition. The resistivity began to decrease at 9.0 and 11.0 K, which corresponds to the onset temperature ($T_c^{onset}$), and it reached zero at 8.8 and 9.2 K ($T_c^{zero}$) for #1 and #2 under 0 T. $T_c$ decreased with an increase in magnetic field. Figure 3(d) shows the magnetic field-temperature phase diagrams for #1 and #2. To obtain the upper critical field $H_{c2}$ ($T$ = 0 K), the resistive midpoints ($T_c^{\rho}$) are plotted in Figure 3 (d). The $H_{c2}$ is estimated as 10.4 and 13.3 T for #1 and #2, using the Werthamer-Helfand-Hohenberg (WHH) model of $H_{c2}$ ($T$ = 0 K) = -0.69*$T_c$*(d$H_{c2}$/d$T$)$_{T=T_c}$ [36], which is used for superconductors in a dirty limit. To confirm the bulk nature of the observed superconductivity, the specific heat was measured for samples #1 and #2. The temperature dependence of the specific heat under various magnetic fields ($H$ = 0, 2, 3, 6, and 9 T) is summarized in Figures 4(a) and 4(b). A clear jump was observed below 11 K under 0 T, and the transition temperature shifted to a lower temperature with an increase in the magnetic field. Although a small jump still exists at 9 T in sample #2, which is the highest limit of the magnetic field of our PPMS, we used the data above the jump temperature under 9 T to estimate the Sommerfeld coefficient ($\gamma$) and the coefficient for the lattice specific heat contribution ($\beta$). The estimated $\gamma$ and $\beta$ were 18.9 and 0.16 for #1, and 19.3 mJ/mol K$^2$ and 0.19 mJ/mol K$^2$ for #2, respectively. The Debye temperature ($\theta_D$) was estimated to be 365 K for #1 and 348 K for #2, which is close to those reported for the conventional (pure) Nb$_3M$ system [30–32]. Figure 4(b) shows the electronic contributions of the specific heat ($C_{el}$), which was calculated by subtracting the phonon contributions from the total specific heat. The clear jump in $C_{el}/T$ and the decrease in $C_{el}/T$ at low temperatures suggest that both samples are bulk superconductors. Interestingly, the superconducting transition observed from $C_{el}$ was relatively broad compared to pure Nb$_3M$ systems [23, 24, 27]. A broad transition was also observed in an HEA-type $Co_{0.2}Ni_{0.1}Cu_{0.1}Rh_{0.3}Ir_{0.3}Zr_2$ superconductor [19]. Therefore, the commonality in the broad specific heat jump at $T_c$ may indicate the possibility of a universal phenomenon for HEA-type superconductors. The nature of the broad transition may imply a semi-continuous evolution of the superconducting gap caused by the HEA effects. To reveal the origin of this phenomenon, further information from the specific heat analyses by functional parameter fitting or experiments, which can directly observe the superconducting gap structure, such as scanning tunneling microscopy (STM), is required. Although the origin of the broad transition is unclear, the present data are sufficient to confirm the emergence of bulk superconductivity in these samples. Compared to the broader specific heat transition of #2, #1 exhibited a sharper transition. This is consistent with the sharp transition of resistivity and magnetization in sample #1 compared to those of #2. From the temperature dependences of $C_{el}/T$, $\Delta C_{sc}/\gamma T_c$ was estimated as 1.64 for #1 and 1.67 for #2. Because the $\Delta C_{sc}/\gamma T_c$ values are slightly greater than $\Delta C_{sc}/\gamma T_c = 1.43$, which is expected from conventional weak-coupling pairing [37], the present samples would be strong-coupling superconductors similar to the pure $Nb_3M$ samples.

Here, we discuss the effect of HEA on the superconducting states of the obtained samples. The present $T_c$s are approximately half for those of the pure $Nb_3M$ samples [30–32, 38]. The large decade of the $T_c$s for the present samples can be related to the introduction of the short-range disorder of atomic ordering due to the HEA effect. There are two scenarios for the decrease in $T_c$ caused by the disorder. The first scenario is the decrease in the electronic density of state [38–41] and the second is the increase in Coulomb interaction due to electron localization [38, 42]. Considering the smaller $\gamma$ values of the present sample than those for pure $Nb_3M$ (Table 1), the reduction in the electronic density of state can explain the obtained results. In addition, considering the highly disordered (alloyed) situation for the present samples, we consider that the second scenario is also reasonable. In the second scenario, it is known that $T_c$ decreases with increasing Coulomb interaction due to electron localization in a sample with an electrical resistivity of 40 μΩ cm or higher [28, 36], that is, a sample having a large amount of disorder due to defects, impurities, or alloying. This is consistent with the resistivity values of approximately 150 μΩ cm for samples #1 and #2. For previously reported conventional alloying superconductors such as HEA superconductors [3–6], the decrease in $T_c$ is not remarkable because of the disorder due to the long coherence length. In contrast, when the superconductors have a short coherence length such as Nb$_3M$, disorder, which potentially causes nanoscale phase separation and/or the formation of nanoscale domains, the superconducting pairing may be affected and hence result in a decrease in $T_c$.

## 4. Conclusions

Herein, we have reported the synthesis and superconducting properties of new HEA-type superconductors, Nb$_3$Al$_{0.2}$Sn$_{0.2}$Ge$_{0.2}$Ga$_{0.2}$Si$_{0.2}$ and Nb$_3$Al$_{0.3}$Sn$_{0.3}$Ge$_{0.2}$Ga$_{0.1}$Si$_{0.1}$. Polycrystalline samples were prepared using pure metals via arc melting. The composition of the obtained samples satisfied the definition of HEA. Superconducting transitions were observed at 9.0 and 11.0 K for #1 and #2, respectively, through magnetization, electrical resistivity, and specific heat measurements. From the resistivity measurement under magnetic fields, the upper critical fields $H_{c2}(0)$ were estimated as 10.4 and 13.3 T. The bulk nature of superconductivity was confirmed from the specific heat jump at $T_c$. Although the estimated Sommerfeld coefficient was slightly lower than that for pure Nb$_3M$, the Debye temperature and $\Delta C/\gamma T_c$ were close to those reported for pure Nb$_3M$. Even though the observed $T_c$ was almost half of that for pure Nb$_3M$, the discovery of superconductivity in HEA-type Nb$_3M$ provides an additional pathway for exploring novel HEA-type superconductors and investigations on the relationship between the HEA effect and superconductivity in highly disordered compounds.

**Declaration of competing interests**

The authors declare that they have no known competing financial interests or personal relationships that could have appeared to influence the work reported in this paper.

**Acknowledgments**

The authors thank M. Yamashita, M. Katsuno, Y. Goto, and O. Miura for their assistance. This work was partly supported by JSPS-KAKENHI (Grant Nos. 18KK0076) and Tokyo Metropolitan Government Advanced Research (H31-1).

Table 1. Parameters for these present samples and selected A15s with $T_c$ above 17 K. Chemical

| Sample | $\Delta S_{mix}/R$ | $a$ (Å) | $\gamma$ (mJ/mol K$^2$) | $T_c$ (K) | $H_{c2(0)}$ (T) | $\Theta_D$ (K) | $\Delta C/\gamma*T_c$ | Ref. |
|---|---|---|---|---|---|---|---|---|
| Nb$_3$(Al$_{0.2}$Sn$_{0.2}$Ge$_{0.2}$Ga$_{0.2}$Si$_{0.2}$) | 1.59 | 5.183 | 18.9 | 9.0 | 10.4 | 364.8 | 1.64 | This work |
| Nb$_3$(Al$_{0.3}$Sn$_{0.3}$Ge$_{0.2}$Ga$_{0.1}$Si$_{0.1}$) | 1.50 | 5.192 | 19.3 | 11.0 | 13.3 | 347.7 | 1.67 | This work |
| Nb$_3$Al | - | 5.180 | 34.9~45.0 | 18.0 | 33.0 | 296 | 1.59~2.78 | 26,27,30,31,38 |
| Nb$_3$Sn | - | 5.289 | 35~52.4 | 18.0 | 29.0 | 208~290 | 1.30~1.7 | 30,31,38 |
| Nb$_3$Ge | - | 5.125 | 30.3 | 21.8~23.9 | 38.0 | 302 | 1.64 | 30,31,38 |
| Nb$_3$Ga | - | 5.171 | 46 | 19.8~20.2 | 34.1~35 | 280 | 1.7~1.74 | 31,33,38 |
| Nb$_3$Si | - | 5.09 | 24~66.4 | 17.4~18.0 | 14.1~15.5 | 310~352 | 1.7 | 24,31,33,38 |

composition, configurational mixing entropy ($\Delta S_{mix}/R$), lattice constant ($a$), electronic specific heat coefficient ($\gamma$), transition temperature ($T_c$), upper critical fields ($H_{c2(0)}$), Debye temperature ($\Theta_D$), and electron-phonon interaction ($\Delta C/\gamma*T_c$) of A15-type superconductors.

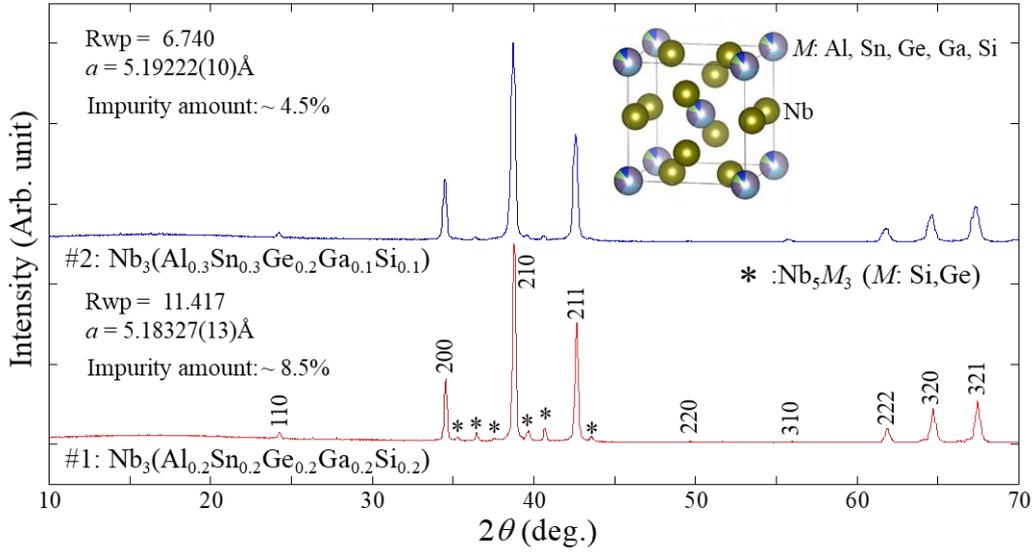

Fig. 1 X-ray diffraction (XRD) patterns for the obtained samples #1 and #2. Inset shows a schematic image of A15-type Nb$_3$M (M = Al, Sn, Ge, Ga, Si).

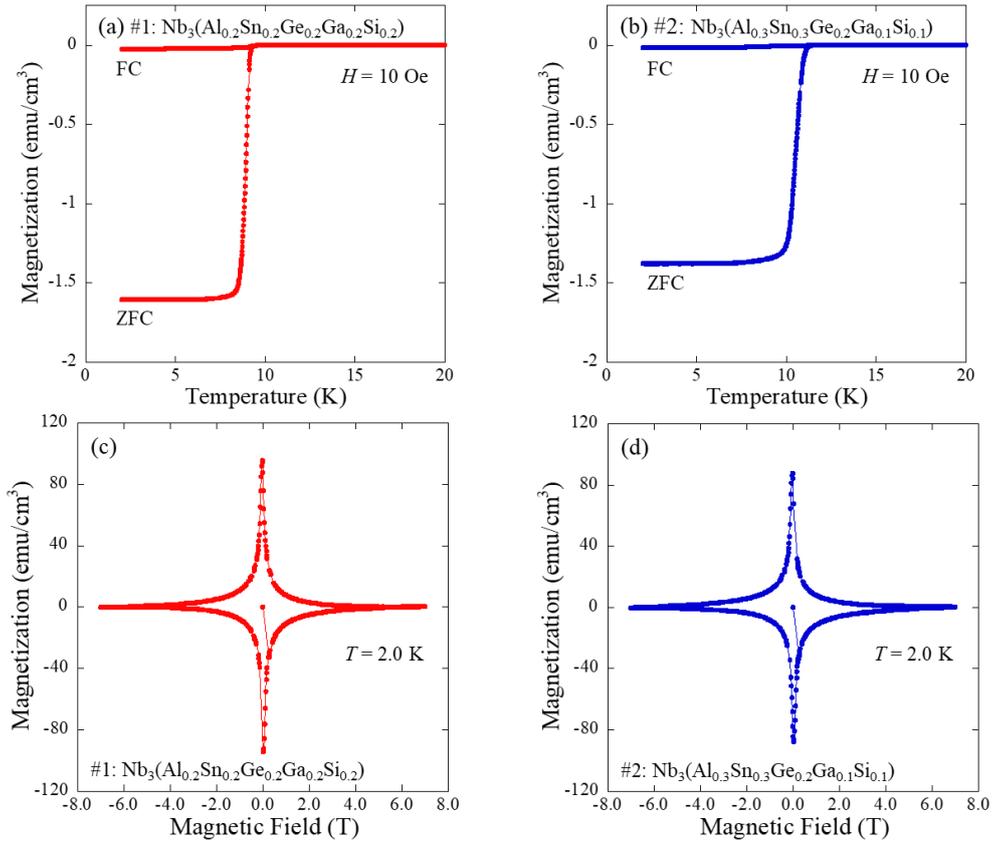

Fig. 2 Temperature dependences of magnetization for samples #1 and #2 (a, b) and M-H- loop at 2.0 K (c, d).

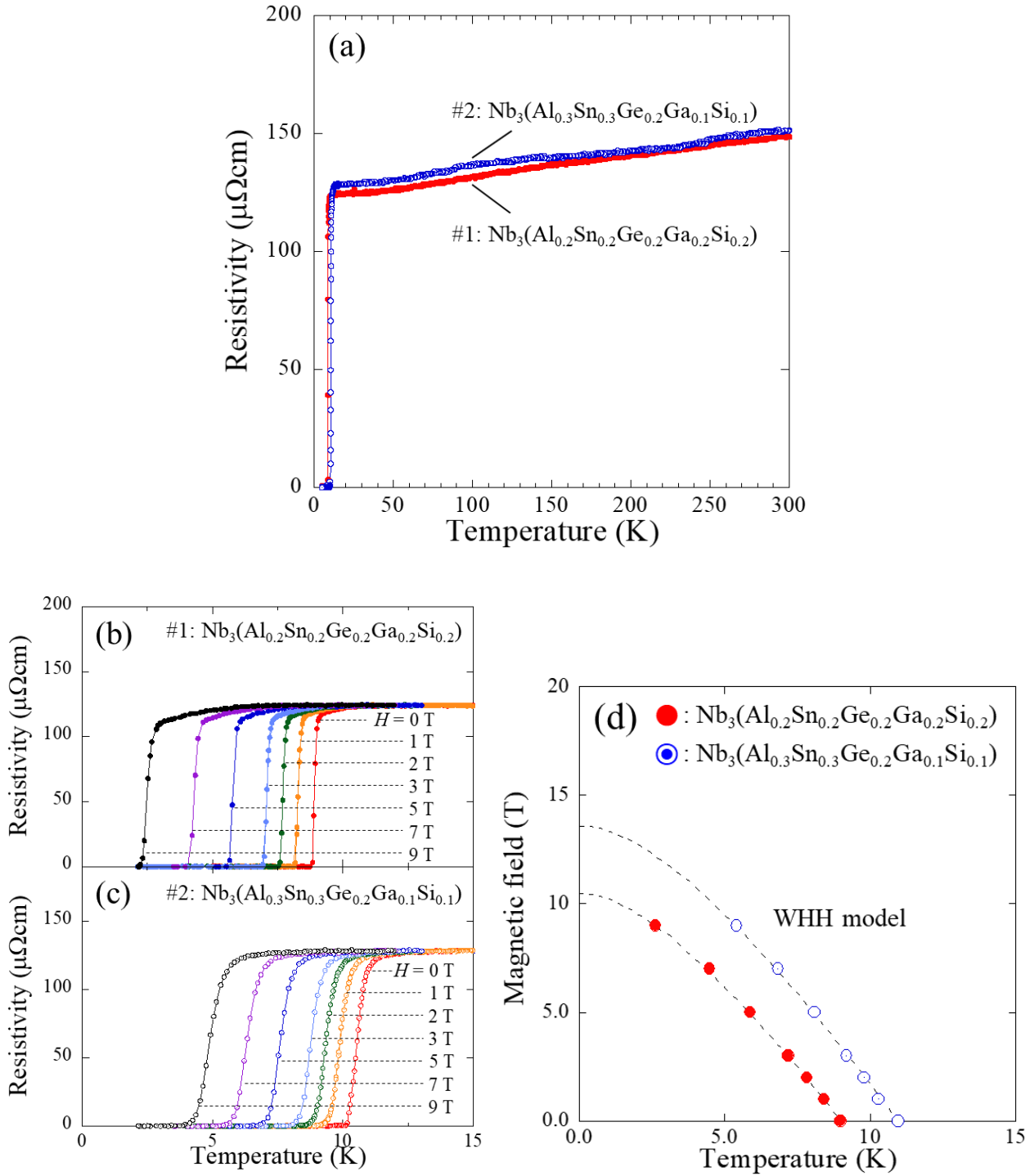

Fig. 3 Temperature dependences of electrical resistivity [$\rho(T)$] for samples #1 and #2 (a). Enlarged $\rho(T)$ to 0 K for samples #1 and #2 under the magnetic fields of 0-9 T (b, c). Temperature dependence of $H_{c2}$ for #1 and #2 where $T_c$ was estimated as the temperature at the resistive midpoint ($T_c^\rho$) (d). The dashed lines are the Werthamer-Helfand-Hohenberg (WHH) fitting results.

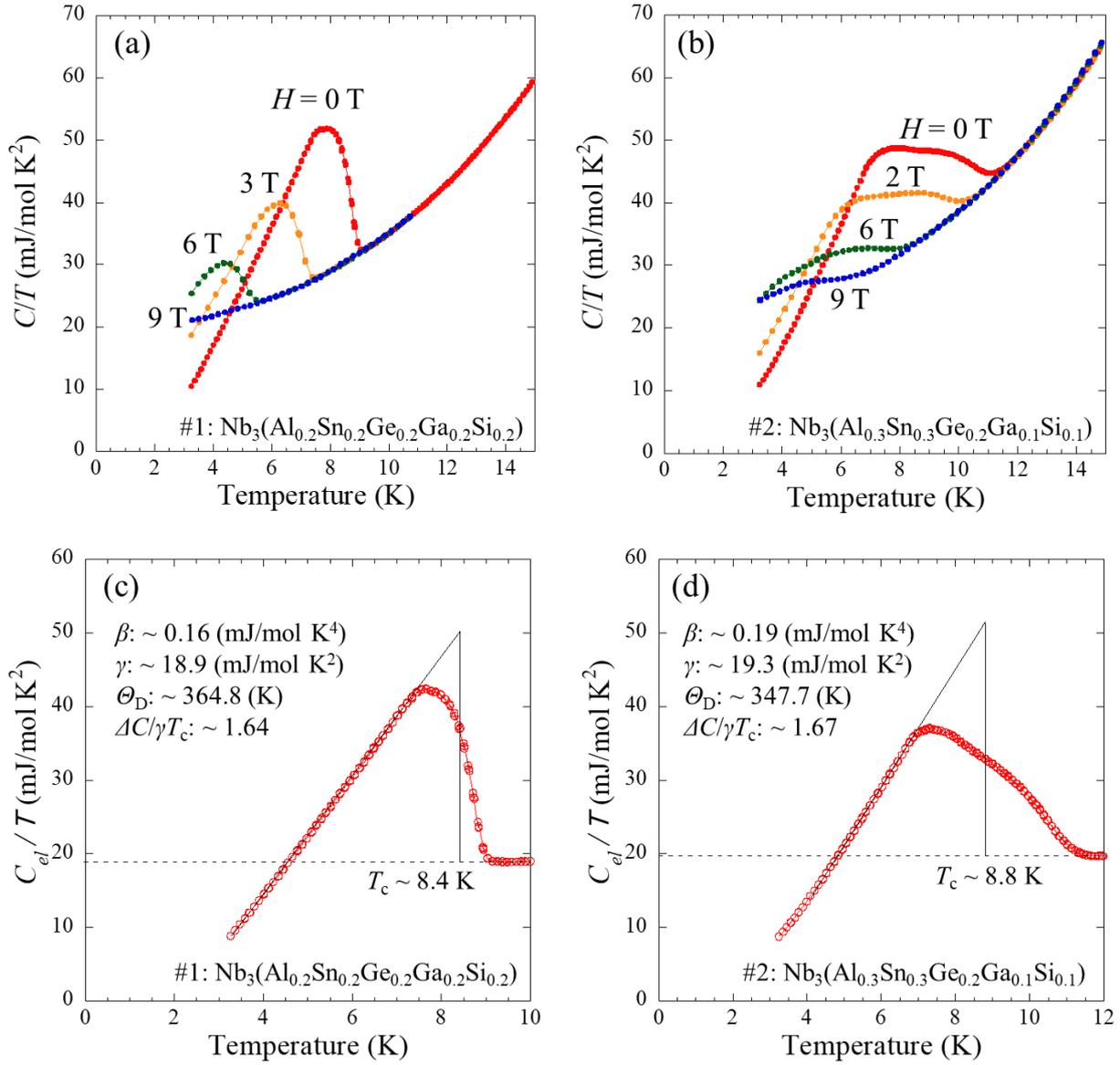

Fig. 4 Temperature dependence of $C/T$ for #1 $Nb_3Al_{0.2}Sn_{0.2}Ge_{0.2}Ga_{0.2}Si_{0.2}$ at 0, 3, 6 and 9 T and #2 $Nb_3Al_{0.3}Sn_{0.3}Ge_{0.2}Ga_{0.1}Si_{0.1}$ at 0, 2, 6 and 9 T (a, b). Temperature dependence of electronic specific heat $C_{el}/T$ for #1 and #2 at 0 T (c, d).